\documentclass[12pt]{iopart}

\usepackage{iopams}  
\usepackage{graphicx}
\usepackage{dcolumn}
\usepackage{bm}
\usepackage{braket}
\usepackage{color}

\begin{document}

\title[Interferometric extraction of photoionization-path amplitudes and phases]{Interferometric extraction of photoionization-path amplitudes and phases from time-dependent multiconfiguration self-consistent-field simulations}

\author{Yuki Orimo$^1$, Oyunbileg Tugs$^1$, Takeshi Sato$^{1,2,3}$, Daehyun You$^4$, Kiyoshi Ueda$^4$ and Kenichi L.~Ishikawa$^{1,2,3}$}

\address{$^1$ Department of Nuclear Engineering and Management, Graduate School of Engineering, The University of Tokyo, 7-3-1 Hongo, Bunkyo-ku, Tokyo 113-8656, Japan}
\address{$^2$ Photon Science Center, Graduate School of Engineering, The University of Tokyo, 7-3-1 Hongo, Bunkyo-ku, Tokyo 113-8656, Japan}
\address{$^3$ Research Institute for Photon Science and Laser Technology, The University of Tokyo, 7-3-1 Hongo, Bunkyo-ku, Tokyo 113-0033, Japan}
\address{$^4$ Institute of Multidisciplinary Research for Advanced Materials, Tohoku University, Sendai 980-8577, Japan}
\ead{ykormhk@atto.t.u-tokyo.ac.jp}
\ead{ishiken@n.t.u-tokyo.ac.jp}

\begin{abstract}
Bichromatic extreme-ultraviolet pulses from a seeded free-electron laser enable us to measure photoelectron angular distribution (PAD) as a function of the relative phase between the different wavelength components. 
The time-dependent multiconfiguration self-consistent-field (TD-MCSCF) methods are powerful multielectron computation methods to accurately simulate such photoionization dynamics from the first principles.
Here we propose a method to evaluate the amplitude and phase of each ionization path, which completely determines the photoionization processes, using TD-MCSCF simulation results. 
The idea is to exploit the capability of TD-MCSCF to calculate the partial wave amplitudes specified by the azimuthal and magnetic angular momenta $(l,m)$ and the $m$-resolved PAD.
The phases of the ionization paths as well as the amplitudes of the paths resulting in the same $(l,m)$ are obtained through global fitting of the expression of the asymmetry parameters to the calculated $m$-resolved PAD, which depends on the relative phase of the bichromatic field.
We apply the present method to ionization of Ne by combined fundamental and second-harmonic XUV pulses, demonstrating that the extracted amplitudes and phases excellently reproduce the asymmetry parameters.
\end{abstract}

%
%
%
%
%

\section{Introduction}

Coherent optics experiments have recently become possible in the extreme-ultraviolet (XUV) spectral range with a seeded free-electron laser (FEL) such as FERMI, and temporally coherent, multi-harmonic XUV pulses with a controllable phase relationship \cite{Diviacco2011} are used to perform coherent control experiments ~\cite{Prince16, SciReport, Callegari2020PRarXiv, DiFraia2019, Iablonskyi}.
These experiments typically measure photoelectron angular distribution (PAD) as a function of the relative phase between different wavelength components.
Quantum mechanical processes including photoionization are described by (real-valued) amplitudes and phases of the paths involved, and the determination of all of these are called complete experiments~\cite{Beckerlanger, Kleinpoppen, Carpeggiani2018}. 
Although amplitudes can often be deduced from measured or calculated signal intensities, the phases are usually more challenging to determine. 

To be specific, as in the experiment carried out at FERMI \cite{DiFraia2019, You2020PRX}, let us consider an atom interacting with a collinearly polarized $\omega$-$2\omega$ two-color pulse, whose field is given by,
\begin{equation}
    \label{eq:field}
	E(t) = \sqrt{I_\omega (t)} \cos \omega t + \sqrt{I_{2\omega} (t)} \cos (2\omega t - \phi),
\end{equation}
where $I_\omega (t)$ and $I_{2\omega} (t)$ denote the envelopes of the two pulses, and $\phi$ the $\omega-2\omega$ relative phase.
The interference is created in photoemission between two-photon ionization (TPI) by a fundamental component ($\hbar\omega$) and single-photon ionization (SPI) by its second harmonic ($2\hbar\omega$).
The photoelectron angular distribution (PAD) $I(\theta)$ is expressed as,
\begin{equation}
\label{eq:PAD}
	I(\theta) \propto 1+\sum_{l=1}^{4}\beta_l P_l(\cos\theta),
\end{equation}
with the Legendre polynomials $P_l(\cos\theta)$. 
The corresponding asymmetry parameters $\beta_l\,(l=1,\cdots,4)$ are, in turn, expressed in terms of the amplitudes and phases of the ionization paths (see below). 
The odd-order $\beta_1$ and $\beta_3$ sinusoidally oscillate with $\phi$, while the even-order $\beta_2$ and $\beta_4$ are constant.

In the simple situation of photoionization of the $s$ electron, typically in He \cite{DiFraia2019},
single-photon ionization leads to a $p$ photoelectron, while two-photon ionization leads to two final continuum states $s$ and $d$.
Thus, we have two unknown amplitude ratios and two unknown phase differences, and we can determine them using the four asymmetry parameters obtained from PAD measurements, as has been done in Ref.~\cite{DiFraia2019}.
We now turn to photoemission of a $p$ electron, e.g., in the other rare gas atoms (Ne \cite{You2020PRX}, Ar, $\cdots$).
There are more ionization pathways: two SPI pathways ($p\to s$ and $p\to d$), together with three TPI pathways ($p\to s \to p$, $p\to d \to p$ and $p \to d \to f$).
We cannot determine the amplitude ratios and phase differences from the four asymmetry parameters, 
even if we note that the amplitudes for different magnetic quantum numbers $m (=0,\pm 1)$ are mutually related by the Wigner-Eckart theorem and that the phases do not depend on $m$.

A class of powerful real-time {\it ab initio} approaches to compute nonlinear laser-atom interactions as considered here are the time-dependent multiconfiguration self-consistent field (TD-MCSCF) methods \cite{Ishikawa_2015,Zanghellini_2003,Kato_2004,Caillat_2005,Nguyen-Dang_2007,Miyagi_2013,Sato2013,Miyagi_2014,Haxton_2015,Sato2015PRA,Sato2016}.
The TD-MCSCF methods express the total electronic wave function in the multiconfiguration expansion,
\begin{equation}
	\label{eq:TD-MCSCF}
	\Psi (t) = \sum_{I} C_I(t) \Phi_I(t),
\end{equation}
where the electronic configuration $\Phi_I(t)$ is a Slater determinant composed of spin orbital functions.
Not only the configuration-interaction coefficients $\{C_I\}$ but also orbitals are evolved in time, in order to describe excitation and ionization efficiently.
We have developed variants of the TD-MCSCF methods called the time-dependent complete-active-space self-consistent field (TD-CASSCF) \cite{Sato2013,Sato2016} and the time-dependent occupation-restricted multiple-active-space (TD-ORMAS) method \cite{Sato2015PRA}, to which we have recently implemented infinite-range exterior complex scaling (irECS) \cite{Orimo2018, PhysRevA.81.053845} as an efficient absorbing boundary and the time-dependent surface flux (tSURFF) method \cite{Tao2012,Orimo2019} to calculate the angle-resolved photoelectron energy spectrum (ARPES).
The excellent agreement with the experimental results of Ne photoionization by bichromatic FEL pulses \cite{You2020PRX} demonstrates high numerical accuracy of the methods.

It is not straightforward to obtain the phase of each ionization pathway or photoelectron partial wave from TD-MCSCF simulation results.
This is because every orbital changes with time and, in principle, becomes partially ionized. 
Hence, it is not trivial to decompose the wave function expanded as equation (\ref{eq:TD-MCSCF}) into a departing photoelectron and the ionic core.
Moreover, the ionized part of each orbital is absorbed at the simulation boundary to prevent unphysical reflection of photoelectron wave packets.

In the present article, we propose a new method to extract the amplitudes and phases of different SPI and TPI pathways from TD-MCSCF simulation results. 
The idea is to analyze the contribution from $m=0$ and $m=\pm 1$ separately. 
Since the PAD $I_m (\theta)$ for each $m$ contains the asymmetry parameters up to $\beta_6$, we gain substantially more information than from the PAD $I(\theta)$ summed over $m$.
Furthermore, the use of global fitting with the bichromatic, coherent-control setup equation (\ref{eq:field}) removes the ambiguity in phase determination arising from the 2-valuedness of arccos within a $2\pi$ interval.
The numerical assessment using the TD-CASSCF method shows that the extracted amplitudes and phases reproduce the original asymmetry parameters very well, validating our method.

This paper is organized as follows. We briefly review the TD-CASSCF method in section \ref{sec:TD-CASSCF} and the tSURFF method for the calculation of ARPES in section \ref{sec:tSURFF}. 
In section \ref{sec:extraction}, we present the extraction of the amplitudes and phases of photoionization pathways, along with numerical results and discussions.
Conclusions are given in section \ref{sec:conclusions}. 
We use Hartree atomic units unless otherwise stated.

\section{TD-CASSCF Method} 
\label{sec:TD-CASSCF}

We consider an $N$-electron atom (or ion) with atomic number $Z$ irradiated by a laser field ${\bf E}(t)$ linearly polarized along the $z$ axis.
In the velocity gauge and within the dipole approximation, its dynamics is described by the time-dependent Schr\"{o}dinger equation (TDSE),
\begin{equation}
\label{eq:TDSE}
i\frac{\partial\Psi (t)}{\partial t} = \hat{H}(t)\Psi (t),
\end{equation}
where the time-dependent Hamiltonian is
\begin{equation}
\hat{H}(t)=\hat{H}_1(t)+\hat{H}_2,
\end{equation}
with the one-electron part
\begin{equation}
\label{eq:H1}
\hat{H}_1(t) = \sum_{i=1}^{N} \hat{h}({\bf r}_i,\nabla_i,t) 
\end{equation}
and the two-electron part
\begin{equation}
\hat{H}_2 = \sum_{i=1}^{Z} \sum_{j > i} \frac{1}{|{\bf r}_i - {\bf r}_j|}.
\end{equation}
The one-body Hamiltonian is given by,
\begin{equation}
\label{eq:velocity-gauge}
\hat{h}({\bf r},\nabla,t) = -\frac{ \nabla^2}{2} - \frac{Z}{|{\bf r}_i|} - i{\bf A}(t) \cdot \nabla,
\end{equation}
where ${\bf A}(t) = -\int {\bf E}(t)dt$ is the vector potential.

In the TD-CASSCF method, the total wave function is given by,
\begin{equation}
	\Psi (t) = \hat{A}\left[\Phi_{\rm fc}\Phi_{\rm dc}(t)\sum_I\Phi_I(t) C_I(t)\right],
\end{equation}
where $\hat{A}$ denotes the antisymmetrization operator, $\Phi_{\rm fc}$ and $\Phi_{\rm dc}$ the closed-shell determinants formed with numbers $n_{\rm fc}$ frozen-core and $n_{\rm dc}$ dynamical-core orbitals, respectively, and $\{\Phi_I\}$ the determinants constructed from  $n_{\rm a}$ active orbitals. 
The active electrons are fully correlated among the active orbitals.
Thanks to this decomposition, we can reduce the computational cost without sacrificing the accuracy in the description of correlated multielectron dynamics.

The equations of motion (EOMs) that describe the temporal evolution of the CI coefficients $\{C_{I}\}$ and the orbital functions $\{\psi_p\}$ are derived by use of the time-dependent variational principle \cite{Sato2015PRA,Moccia_1973}.
and read,
\begin{equation}
    \label{eomci}
    i \frac{d}{dt} C_I(t) = \sum_J \braket{\Phi_I| \hat{H} - \hat{R}| \Phi_J}
\end{equation}
\begin{equation}
    \label{eomorb}
	i \frac{d}{dt} \ket{\psi_p} = \hat{h} \ket{\psi_p} + \hat{Q} \hat{F}  \ket{\psi_p} + \sum_{q} \ket{\psi_q}  R^q_p,	
\end{equation}
where $\hat{Q} = 1 - \sum_{q} \Ket{\psi_q} \Bra{\psi_q}$ the projector onto the orthogonal complement of the occupied orbital space.
$\hat{F}$ is a non-local operator describing the contribution from the interelectronic Coulomb interaction, defined as
\begin{equation}
	\hat{F} \ket{\psi_p} = \sum_{oqsr} (D^{-1})^o_p P^{qs}_{or} \hat{W}^r_s \ket{\psi_q},
\end{equation}
where $D$ and $P$ are the one- and two-electron reduced density matrices, and $\hat{W}^r_s$ is given, in the coordinate space, by
\begin{equation}
	W^{r}_{s} \left(\boldsymbol{r} \right) = \int d \boldsymbol{r}^\prime \frac{\psi_{r}^{*} (\boldsymbol{r}^\prime) \psi_{s} ( \boldsymbol{r}^\prime )}{| \boldsymbol{r} - \boldsymbol{r}^\prime | } .
	\label{eq:W}
\end{equation}
The matrix element $R^q_p$ is given by,
\begin{equation}
\label{eq:orbital-time-derivative}
	R^q_p = i \braket{\psi_q | \dot{\psi_p} } - h^q_p,
\end{equation}
with $h^q_p = \braket{\psi_q|\hat{h}|\psi_p}$. 
$R^q_p$'s within one orbital subspace (frozen core, dynamical core and each subdivided active space) can be arbitrary Hermitian matrix elements, and in this paper, they are set to zero. 
On the other hand, the elements between different orbital subspaces are determined by the TDVP. Their concrete expressions are given in Ref.~\cite{Sato2015PRA}, where $iX^q_p = R^p_q + h^q_p$ is used for working variables.

Our numerical implementation \cite{Sato2016} employs a spherical harmonics expansion of orbitals with the radial coordinate discretized by a finite-element discrete variable representation (FEDVR) \cite{PhysRevA.62.032706,McCurdy_2004,PhysRevE.73.036708,quant_dyn_imag}.
Specifically, we do TD-CASSCF calculations where each orbital is expanded with spherical harmonics whose largest angular momentum is 6.
The radial coordinate spanning up to 44 a.u. is divided into 11 finite elements each of which contains 23 DVR grid points and an absorbing boundary using infinite-range exterior complex scaling (irECS) \cite{Orimo2018,PhysRevA.81.053845} is placed at 40 a.u. with one additional finite element extending to infinity. 
The time step is 1/600 of an optical cycle of the $\omega$ pulse.
The initial ground state is obtained through imaginary time propagation of the EOMs. 

We perform TDHF and TD-CASSCF simulations for Ne. Since the pulse is non-resonant, the PAD is insensitive to the pulse width. Table \ref{table:pulse-parameters} lists $\omega$ and peak intensities used. We use one frozen-core and eight active orbitals in the TD-CASSCF calculations.
Note that we use two different $2\omega$ intensities for $\hbar\omega = 15.9\,{\rm eV}$.
The full-width-at-half-maximum pulse length is chosen to be 10 fs. It has been shown that the pulse length does not affect the result, provided the photoionization is non-resonant, i.e. no resonances occur within the photon bandwidth \cite{Ishikawa2012, Ishikawa2013, You2020PRX}.

\begin{table}[tb]
\begin{center}
\caption{\label{table:pulse-parameters} Photon energy and peak intensity of the pulse used in our simulations.}
\begin{tabular}{ccccc}
\br

    Label & $\hbar\omega$ (eV) & $I_\omega ({\rm W/cm}^2)$ & $I_{2\omega} ({\rm W/cm}^2)$ \\
\mr
    A & 14.3 & $10^{13}$ & $2.32 \times 10^{8}$  \\
    B & 15.9 & $10^{13}$ & $1.18 \times 10^{8}$  \\
    C & 15.9 & $10^{13}$ & $4.21 \times 10^{8}$  \\
    D & 19.1 & $10^{13}$ & $2.82 \times 10^{8}$  \\
\br
\end{tabular}
\end{center}
\end{table}

\section{Photoelectron angular distribution}
\label{sec:tSURFF}

From the obtained time-dependent wave functions, we extract the angle-resolved photoelectron energy spectrum (ARPES) by use of the time-dependent surface flux (tSURFF) method \cite{Tao2012,Orimo2019}. 
This method computes the ARPES from the electron flux through a surface located at a certain radius $R_s$, beyond which the outgoing flux is absorbed by the infinite-range exterior complex scaling \cite{Orimo2018,PhysRevA.81.053845}.

We introduce the time-dependent momentum amplitude $a_p({\bf k},t)$ of orbital $p$ for photoelectron momentum ${\bf k}$, defined by
\begin{equation}
	a_p({\bf k},t) =  \langle\chi_{\bf k}({\bf r},t)|u(R_s)|\psi_p ({\bf r},t)\rangle
    \equiv \int_{r>R_s}\chi_{{\bf k}}^*({\bf r},t)\psi_p({\bf r},t)d^3 {\bf r},
\end{equation}
where $\chi_{{\bf k}}({\bf r},t)$ denotes the Volkov wavefunction, and $u(R_s)$ the Heaviside function which is unity for $r>R_s$ and vanishes otherwise. The use of the Volkov wavefunction implies that we neglect the effects of the Coulomb force from the nucleus and the other electrons on the photoelectron dynamics outside $R_s$, which has been confirmed to be a good approximation \cite{Orimo2019}.
The photoelectron momentum distribution $\rho ({\bf k})$ is given by
\begin{equation}
	\label{eq:PEMD}
	\rho({\bf k}) 
	=  \sum_{pq} a_p({\bf k},\infty) a_q^*({\bf k},\infty) D^p_q,
\end{equation}
%
One obtains $a_p({\bf k},\infty)$ by numerically integrating,
\begin{eqnarray}
	-i \frac{\partial}{\partial t} a_p({\bf k},t) &=& \langle\chi_{{\bf k}}(t)|[\hat{h}_s, u(R_{s})]|\psi_p(t)\rangle \nonumber\\
	&+& \sum_q a_q({\bf k},t) \left[\langle\psi_q(t)|\hat{F}|\psi_p(t)\rangle - R^q_p\right],\label{eomme_tsurff}
\end{eqnarray}
where $\hat{h}_s$ denotes the Volkov Hamiltonian,
\begin{equation}
	\hat{h}_s=-\frac{ \nabla^2}{2} - i{\bf A}(t) \cdot \nabla.
\end{equation}
For the case of linear polarization along the $z$ axis, the magnetic quantum number of each orbital is conserved, and $D^p_q$ vanishes if orbitals $p$ and $q$ have different magnetic quantum numbers $m_p\ne m_q$ \cite{Sato2016}.
Hence, $\rho ({\rm k})$ can be decomposed into the momentum distribution of photoelectrons with different magnetic quantum numbers $m$ as,
\begin{equation}
	\label{eq:PEMDm}
	\rho ({\bf k}) = \sum_m \rho_m ({\bf k}), \quad \rho_m ({\bf k})=\sum_{pq\, (m_p=m_q=m)} a_p({\bf k},\infty) a_q^*({\bf k},\infty) D^p_q
\end{equation}
The numerical implementation of tSURFF to TD-MCSCF is detailed in Ref.~\cite{Orimo2019}.
We evaluate the photoelectron angular distribution $I \left( \theta;\phi \right)$ as a slice of $\rho \left( {\bf k} \right)$
at the value of $\left| {\bf k} \right|$ corresponding to the photoelectron peak.

\section{Extraction of amplitudes and phases from the numerical results} 
\label{sec:extraction}

The PAD $I(\theta)$ is expressed as \cite{Amusia1990,You2020PRX},
\begin{eqnarray}
\label{eq:NePAD}
I(\theta)&=& \sum_{m=-1}^{1} I_m(\theta) \\
&=&\left|c_{pd}^{-1} e^{i \eta_{pd}} \, Y_{1}^{-1}\left(\theta, \varphi \right) +
c_{d}^{-1} e^{i \left( \eta_{d} + \phi \right)} \, Y_{2}^{-1}\left(\theta, \varphi \right)+
c_{fd}^{-1} e^{i \eta_{fd}} \, Y_{3}^{-1}\left(\theta, \varphi \right) \right|^2 \nonumber\\
&+&
|c_{s}^0 e^{i \left( \eta_{s} + \phi \right)} \, Y_{0}^{0}\left(\theta, \varphi \right) +
c_{ps}^0 e^{i \eta_{ps}} \, Y_{1}^{0}\left(\theta, \varphi \right) +
c_{pd}^0 e^{i \eta_{pd}} \, Y_{1}^{0}\left(\theta, \varphi \right) \nonumber\\
&+&c_{d}^0 e^{i \left( \eta_{d} + \phi \right)} \, Y_{2}^{0}\left(\theta, \varphi \right)+
c_{fd}^0 e^{i \eta_{fd}} \, Y_{3}^{0}\left(\theta, \varphi \right)|^2 \nonumber\\
\label{eq:NePADdecomposition} 
&+&
\left| c_{pd}^1 e^{i \eta_{pd}} \, Y_{1}^{1}\left(\theta, \varphi \right) +
c_{d}^1 e^{i \left( \eta_{d} + \phi \right)} \, Y_{2}^{1}\left(\theta, \varphi \right)+
c_{fd}^1 e^{i \eta_{fd}} \, Y_{3}^{-1}\left(\theta, \varphi \right)\right|^2 \\
&=& \frac{B}{4\pi}\left[1+\sum _{ l=1 }^{ 4 }{ \beta _{ l } } { P }_{ l }(\cos { { \theta  }} )\right],
\end{eqnarray}
using outgoing partial wave amplitudes $c_\xi^m$'s (taken to be real) and phases $\eta_\xi$'s, where the subscript $\xi$ denotes the ionization path; $\xi=d$ and $pd$, for example, correspond to SPI $2p\to d$ and TPI $2p\to d \to p$, respectively.
With $I(\theta)$ at hand, either experimentally or computationally, we can calculate the asymmetry parameters $\{\beta_l\}$ by projection as,
\begin{equation}
	\beta_l = \frac{\beta_l^\prime}{\beta_0^\prime},\quad \beta_l^\prime=\frac{2l+1}{2}\int_{-\pi}^\pi I(\theta)P_l(\cos\theta)\sin\theta d\theta,\quad B=4\pi\beta_0^\prime.
\end{equation}
It follows from the Wigner-Eckart theorem that, 
\begin{equation}
	\label{eq:Wigner-Eckart}
    c_\xi^1=c_\xi^{-1},\quad c_{pd}^{\pm 1} = \frac{3}{4}c_{pd}^0,\quad c_d^{\pm 1} = \frac{\sqrt{3}}{2}c_d^0, \quad c_{fd}^{\pm 1}=\frac{\sqrt{6}}{3}c_{fd}^0.
\end{equation}
Then, the asymmetry parameters are given in terms of $c_\xi^0$'s and $\eta_\xi$'s by,
\begin{eqnarray}
	B &=& (c_s^0)^2 + \frac{5}{2}(c_d^0)^2 + \frac{7}{3}(c_{fd}^0)^2 \nonumber\\
	    &+& \frac{17}{8}(c_{pd}^0)^2 + 2 c_{{pd}}^0 c_{{ps}}^0 \cos \left(\eta _{{pd}}-\eta _{{ps}}\right)+(c_{{ps}}^0)^2,
\end{eqnarray}
\begin{eqnarray}
    \label{eq:beta1}
	\beta_1&=& \frac{1}{B}\left[6\sqrt{\frac{7}{5}} c^{0}_{d}c_{fd}^0 \cos{\left (\eta_{d} - \eta_{fd} + \phi \right )}\right. \\
    &+&\frac{17}{2} \sqrt{\frac{3}{5}} c_d^0 c_{{pd}}^0 \cos \left(\eta_d-\eta_{{pd}}+\phi\right)+4 \sqrt{\frac{3}{5}} c_d^0 c_{{ps}}^0 \cos \left(\eta _d-\eta_{{ps}}+\phi\right) \nonumber \\  
    &+& \left. 2 \sqrt{3} c_s^0 c_{pd}^0 \cos \left(\eta_s-\eta_{pd}+\phi\right)+2 \sqrt{3} c_s^0 c_{ps}^0 \cos \left(\eta _s-\eta_{ps}+\phi\right)\right],
\end{eqnarray}
\begin{eqnarray}
    \label{eq:beta2}
    \beta_2 &=& \frac{1}{B}\left[{\frac{5}{2}(c_d^0)^2} + 4\sqrt{5} c_d^0 c_s^0 \cos \left(\eta _d-\eta _s\right)\right. \nonumber \\
    &+& \frac{8}{3}(c_{fd}^0)^2 + \frac{12\sqrt{21}}{7}c_{fd}^0c_{pd}^0\cos{\left (\eta_{fd} - \eta_{pd} \right )} + \frac{7}{8}(c_{pd}^0)^2  \nonumber\\
    &+& \left. 6 \sqrt{\frac{3}{7}} c_{fd}^0 c_{{ps}}^0 \cos \left(\eta _{fd}-\eta _{{ps}}\right)
    + 4 c_{{pd}}^0 c_{{ps}}^0 \cos \left(\eta _{{pd}}-\eta _{{ps}}\right)+2 (c_{{ps}}^0)^2 \right],
\end{eqnarray}
%
%
%
\begin{eqnarray}
    \label{eq:beta3}
    \beta_3 &=&  \frac{1}{B}\left[\frac{4\sqrt{35}}{5} c_d^0 c_{fd}^0 \cos \left(\eta _d-\eta _{fd}+\phi\right)+3 \frac{\sqrt{15}}{10} c_d^0 c_{{pd}}^0 \cos \left(\eta _d-\eta _{{pd}}+\phi\right)\right. \nonumber \\  
    &+& \left. 6 \frac{\sqrt{15}}{5} c_d^0 c_{{ps}}^0 \cos \left(\eta _d-\eta _{{ps}}+\phi\right)+2 \sqrt{7} c_s^0 c_{fd}^0 \cos \left(\eta _s-\eta _{fd}+\phi\right)\right],
\end{eqnarray}
\begin{eqnarray}
    \label{eq:beta4}
    \beta_4 &=& \frac{2}{7B} \left[\sqrt{21}c_{fd}^0c_{{pd}}^0 \cos \left(\eta _{fd}-\eta _{{pd}}\right) + 4 \sqrt{21}c_{fd}^0c_{{ps}}^0 \cos \left(\eta _{fd}-\eta _{{ps}}\right)+7 (c_{fd}^0)^2\right]. \nonumber\\
\end{eqnarray}
$\beta_l$ sinusoidally oscillates with the $\omega$-$2\omega$ relative phase $\phi$ for odd $l$, while it is constant independent of $\phi$ for even $l$.
Also, note that $I_m(\theta) (m=0,\pm 1)$ do not depend on $\varphi$.
Our objective is to obtain the values of $c_\xi^m$'s and $\eta_\xi$'s using the simulation results.
Since we can determine only the phase difference between different paths, we consider $\Delta\eta_\xi\equiv \eta_\xi-\eta_d$ in what follows.

\subsection{TDHF cases}

Let us first consider the case of TDHF simulations,  where the total wave function $\Psi (t)$ is approximated by a single Slater determinant. 
Under the conditions considered in this study, only the three spatial orbitals, one for each $m$, that initially have the $2p$ character will eventually contain the outgoing (ionizing) part. 
Hence, one can obtain the amplitude $c_l^m$ and phase $\eta_l^m$ of each partial wave $(l,m)$ by projecting $a_p({\bf k},\infty)$ for those orbitals onto $Y_{lm}(\theta_k,\varphi_k)$.
It should be noticed that,
\begin{equation}
	c_s^0 = c_0^0, \quad c_{pd}^{\pm 1}=c_1^{\pm 1}, \quad c_d^m=c_2^m, \quad c_{fd}^m=c_3^m,
\end{equation}
\begin{equation} 
	\eta_s = \eta_0^0, \quad \eta_{pd}=\eta_1^{\pm 1}, \quad \eta_d=\eta_2^m, \quad \eta_{fd}=\eta_3^m,
\end{equation}
but that this procedure cannot resolve the paths $2p\to s \to p$ and $2p\to d \to p$ for $m=0$.
Instead, we extract the amplitude $c_1^0$ and phase $\eta_1^0$ that satisfies, 
\begin{equation}
 c_1^0e^{i\eta_1^0}={c_{ps}^0e^{i\eta_{ps}}+c_{pd}^0e^{i\eta_{pd}}}.   
\end{equation}
In this sense, strictly speaking, the projection procedure is not a ``complete experiment".
The results are summarized in Tables \ref{table:amp_tdhf} and \ref{table:scat_tdhf}.
We see that $c_l^1 = c_l^{-1}$ ($c_\xi^1 = c_\xi^{-1}$) and $\eta_{1,3}^1=\eta_{1,3}^{-1}$ are perfectly satisfied, and that $\eta_3^{\pm 1} = \eta_3^0$ is approximately satisfied.

\begin{table}[tb]
\caption{Partial wave amplitudes $c_l^m$calculated with TDHF}
\centering
\label{table:amp_tdhf}

\begin{tabular}{ccccccc}
\br
	Label & $\hbar\omega$ (eV)       &    $m$    & $c_0^m$ & $c_1^m$   & $c_2^m$   & $c_3^m$   \\
\mr
     &       & $-1$ &       & 0.03020 & 0.00981 & 0.03967 \\
A &   14.3    & $0$  & 0.007358 & 0.03767 & 0.01123 & 0.04832 \\
     &          & $1$  &      & 0.03020 & 0.00981 & 0.03967 \\
\mr 
     &           & $-1$ &       & 0.02680 & 0.00638 & 0.03226 \\
B &   15.9    & $0$  & 0.004442 & 0.01240 & 0.00762 & 0.04023 \\
     &          & $1$  &       & 0.02680 & 0.00638 & 0.03226 \\
\mr 
     &            & $-1$ &       & 0.02680 & 0.01202 & 0.03225 \\
C &   15.9   & $0$  & 0.008364 & 0.01239 & 0.01430 & 0.04021 \\
     &          & $1$  &       & 0.02680 & 0.01202 & 0.03225 \\
\mr 
     &           & $-1$ &   & 0.00377 & 0.00787 & 0.02397 \\
D &   19.1      & $0$  & 0.004788 & 0.01648 & 0.00964 & 0.03026 \\
     &          & $1$  &       & 0.00377 & 0.00787 & 0.02397\\
\br
\end{tabular}
\end{table}

\begin{table}[tb]
\caption{Phases $\Delta\eta_l^m=\eta_l^m-\eta_2^m$ calculated with TDHF}
\centering
\label{table:scat_tdhf}
\begin{tabular}{ccccccc}
\br
        Label & $\hbar\omega$ (eV)       &    $m$    & $\Delta\eta_0^m$ & $\Delta\eta_1^m$   & $\Delta\eta_3^m$   \\
\mr
     &       & $-1$ &        & -2.3280 & 1.1622\\
A &   14.3    & $0$ & 2.337 & -2.1519 & 1.1621 \\
      &         & $1$  &      & -2.3280 & 1.1622 \\
\mr 
     &           & $-1$ &        & -2.2765 & 1.2034 \\
B &   15.9    & $0$  & 2.109 & 3.0374       & 1.2060 \\
     &          & $1$  &       & -2.2765      & 1.2034  \\
\mr 
     &            & $-1$ &       & -2.2754      & 1.2040 \\
C &   15.9   & $0$  & 2.111 & 3.0380       & 1.2056 \\
     &          & $1$  &       & -2.2754      & 1.2040\\
\mr 
     &           & $-1$ &   &  2.4571       & 1.2578 \\
D &   19.1      & $0$  &  1.800 & -2.2268      & 1.2658 \\
     &          & $1$  &       & 2.4571       & 1.2578\\
\br
\end{tabular}
\end{table}

\begin{table}[tb]
\caption{Partial wave amplitudes $c_l^m$ calculated from TD-CASSCF}
\centering
\label{table:amp_TD-CASSCF}
\begin{tabular}{ccccccc}
\br
	Label & $\hbar\omega$ (eV)       &    $m$    & $c_0^m$ & $c_1^m$   & $c_2^m$   & $c_3^m$   \\
\mr
  &          & $-1$ &      & 0.03051 & 0.00995 & 0.04508 \\
A &   14.3    & $0$  & 0.007548 & 0.04002 & 0.01162 & 0.05476 \\
  &             & $1$  &      & 0.03051 & 0.00995 & 0.04508 \\
\mr 
  &              & $-1$ &      & 0.03175 & 0.00629 & 0.03709 \\
B &   15.9    & $0$ &0.004326 & 0.01347 & 0.00731 & 0.04450 \\
  &             & $1$  &       & 0.03175 & 0.00629 & 0.03709 \\
\mr 
  &              & $-1$ &         & 0.03173 & 0.01191 & 0.03711 \\
C &   15.9   & $0$ & 0.008174 & 0.01349 & 0.01398 & 0.04452 \\
  &             & $1$ &         & 0.03173 & 0.01191 & 0.03711 \\  
\mr 
  &              & $-1$ &         & 0.00494 & 0.00875 & 0.02688 \\
D &   19.1        & $0$ & 0.005034 & 0.00811 & 0.01012 & 0.03318 \\
  &             & $1$ &         & 0.00494 & 0.00875 & 0.02688\\
\mr
\end{tabular}
\end{table}

\subsection{TD-MCSCF cases}

For the case of more-than-one configurations such as TD-CASSCF, since the total wave functions are highly correlated, we adopt the procedure described in this subsection.

\subsubsection{Partial wave amplitudes.}
\label{subsubsec:Partial wave amplitudes}

To extract the (real-valued) partial wave amplitude $c_l^m$, let us expand $a_p({\bf k},\infty)$ as,
\begin{equation}
\label{eq:apgY}
	a_p({\bf k},\infty) = \sum_l g_p^{l,m_p}(k)Y_{l}^{m_p}(\theta_k,\varphi_k),
\end{equation}
with
\begin{equation}
	g_p^{l,m_p}(k)=\int a_p({\bf k},\infty) Y_{l}^{m_p}(\theta_k,\varphi_k)^*d\Omega_k.
\end{equation}
Then, integrating $\rho_m ({\bf k})$ in equation (\ref{eq:PEMDm}) over the solid angle, we obtain,
\begin{equation}
	\int \rho_m ({\bf k})d\Omega_k = \sum_l \sum_{pq\, (m_p=m_q=m)}	g_p^{l,m}(k) g_q^{l,m}(k)^*D^p_q,
\end{equation}
which indicates that,
\begin{equation}
\label{eq:partial wave amplitude}
	c_l^m = \sqrt{\sum_{pq\, (m_p=m_q=m)}	g_p^{l,m}(k) g_q^{l,m}(k)^*D^p_q}.
\end{equation}
Thus extracted partial wave amplitudes $c_l^m$ are summarized in Table \ref{table:amp_TD-CASSCF}.
$c_l^1 = c_l^{-1}$ ($c_\xi^1 = c_\xi^{-1}$) is satisfied.
Again, we cannot resolve the paths $2p\to s \to p$ and $2p\to d \to p$ for $m=0$ at this stage.
In principle, one can further generalize equation (\ref{eq:partial wave amplitude}) to relate both the partial wave amplitudes and phases to $\{g_p^{l,m}\}$ and $\{D_q^p\}$ (See \ref{appendix:complex partial wave amplitudes}).
Such an approach, however, cannot resolve multiple paths leading to the same partial wave $(l,m)$.

\subsubsection{Extraction of $\Delta\eta_{fd}$ and $\Delta\eta_{pd}$ from $I_{\pm 1}(\theta)$ by global fitting.}

\begin{figure}[tb]
\begin{center}
\includegraphics[width=10cm]{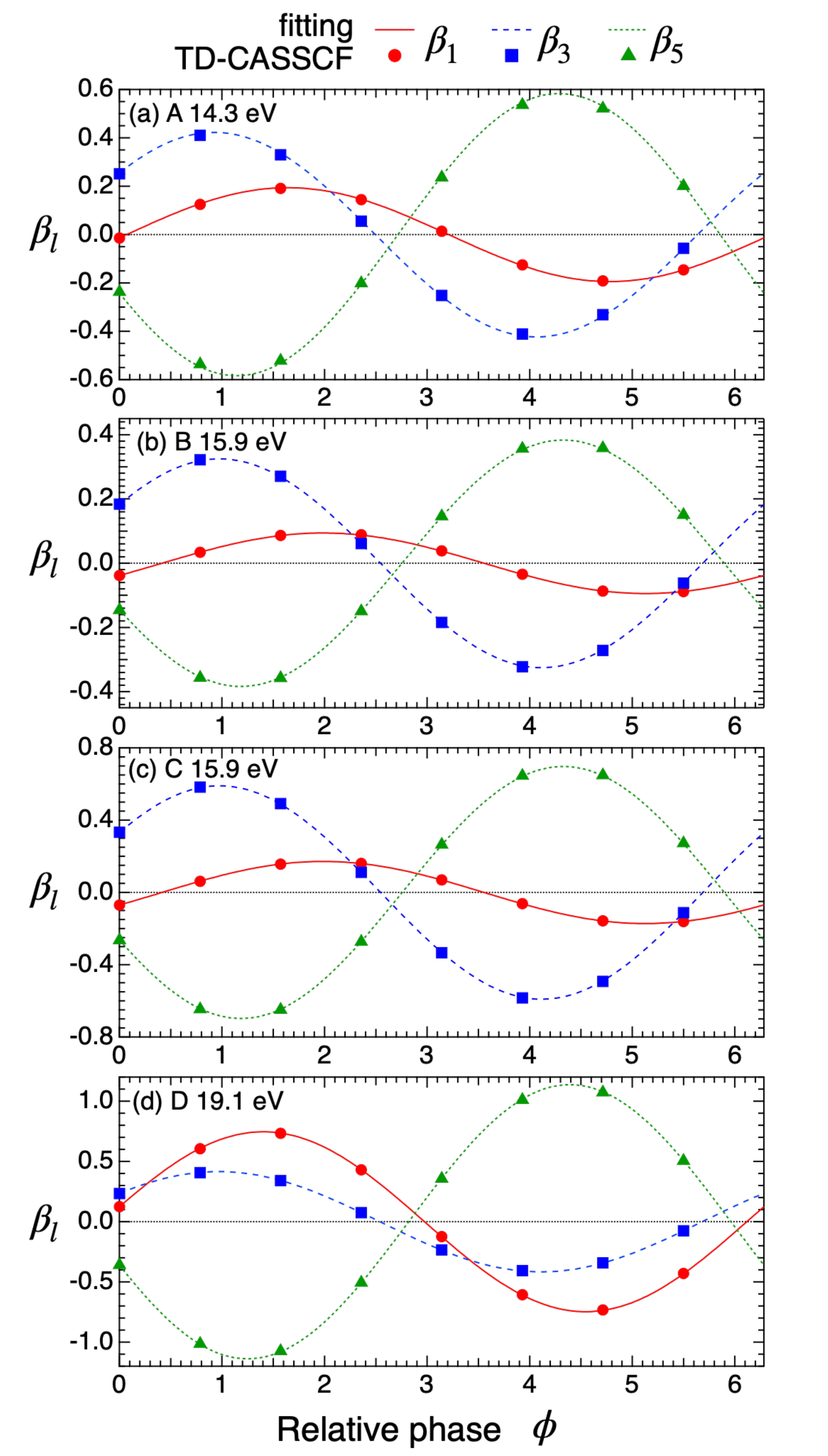}
\caption{Comparison of $\beta_l\,(l=1,3,5)$ as a function of $\phi$ for $m=\pm 1$ between the TD-CASSCF outputs (markers) and the reproduction by equations (\ref{eq:beta1-m-pm1}), (\ref{eq:beta3-m-pm1}), and (\ref{eq:beta5-m-pm1}) using the amplitudes and phases obtained through the global fitting procedure described in the text (lines). The TD-CASSCF calculations were done at eight values of $\phi$ between 0 and $\frac{7}{4}\pi$ with an interval of $\frac{\pi}{4}$.}
\label{fig:m1-fitting}
\end{center}
\end{figure}

Since the amplitudes and phases take the same values for $m=1$ and $-1$, we use the sub- and superscript $\pm 1$.
The PAD $I_{\pm 1}(\theta)$ for $m = \pm 1$ is given by,
\begin{eqnarray}
I_{\pm 1}(\theta)&=&
|{c^{\pm 1}_{pd}} e^{i \eta_{pd}} \, Y_{1}^{\pm 1}\left(\theta, \varphi \right) +
c_{d}^{\pm 1} e^{i \left( \eta_{d} + \phi \right)} \, Y_{2}^{\pm 1}\left(\theta, \varphi \right) \nonumber \\
&+&c_{fd}^{\pm 1} e^{i \eta_{fd}} \, Y_{3}^{\pm 1}\left(\theta, \varphi \right)|^2 \\
\label{eq:PAD(m=pm1)}
&=&\frac{({c^{\pm 1}_{d}})^{2} + ({c^{\pm 1}_{fd}})^{2} + ({c^{\pm 1}_{pd}})^{2}}{4\pi} \left[1+\sum _{ n=1 }^{ 6 }{ \beta _{ n } } { P }_{ n }(\cos { { \theta  }} )\right],
\end{eqnarray}
with, 
\begin{equation}
    \label{eq:beta1-m-pm1}
    \beta_1=\frac{6 c^{\pm1}_{d}\left(2 \sqrt{70} c^{\pm1}_{fd} \cos{\left (\eta_{d} - \eta_{fd} + \phi \right )} + 7 \sqrt{5} c^{\pm1}_{pd} \cos{\left (\eta_{d} - \eta_{pd} + \phi \right )}\right)}{35 \left((c^{\pm1}_{d})^{2} + (c^{\pm1}_{fd})^{2} + (c^{\pm1}_{pd})^{2}\right)} 
\end{equation}
\begin{equation}
   \label{eq:beta2-m-pm1}
    \beta_2= \frac{\Big({5 (c^{\pm1}_{d})^{2}} + 7(c^{\pm1}_{fd})^{2} + {6\sqrt{14} c^{\pm1}_{fd}}  c^{\pm1}_{pd} \cos{\left (\eta_{fd} - \eta_{pd} \right )} - 7(c^{\pm1}_{pd})^{2}\Big)}{7\left((c^{\pm1}_{d})^{2} + (c^{\pm1}_{fd})^{2} + (c^{\pm1}_{pd})^{2}\right)}  
\end{equation}
\begin{equation}
    \label{eq:beta3-m-pm1}
    \beta_3= \frac{2 c^{\pm1}_{d}\left(\sqrt{70} c^{\pm1}_{fd} \cos{\left (\eta_{d} - \eta_{fd} + \phi \right )} - 9 \sqrt{5} c^{\pm1}_{pd} \cos{\left (\eta_{d} - \eta_{pd} + \phi \right )}\right)}{15 \left((c^{\pm1}_{d})^{2} + (c^{\pm1}_{fd})^{2} + (c^{\pm1}_{pd})^{2}\right)} 
\end{equation}
\begin{equation}
    \label{eq:beta4-m-pm1}
    \beta_4= \frac{3\left(- 44 (c^{\pm1}_{d})^{2} + 7 (c^{\pm1}_{fd})^{2} - 22 \sqrt{14} c^{\pm1}_{fd} c^{\pm1}_{pd} \cos{\left (\eta_{fd} - \eta_{pd} \right )}\right)}
    {77 \left((c^{\pm1}_{d})^{2} + (c^{\pm1}_{fd})^{2} + (c^{\pm1}_{pd})^{2}\right)} 
\end{equation}
\begin{equation}
\label{eq:beta5-m-pm1}
    \beta_5= - \frac{10 \sqrt{70} c^{\pm1}_{d} c^{\pm1}_{fd} \cos{\left (\eta_{d} - \eta_{fd} + \phi \right )}} {21 \left((c^{\pm1}_{d})^{2} + (c^{\pm1}_{fd})^{2} + (c^{\pm1}_{pd})^{2}\right)} 
    \end{equation}
\begin{equation}
    \label{eq:beta6-m-pm1}
    \beta_6= - \frac{25 (c^{\pm1}_{fd})^{2}} {11 \left((c^{\pm1}_{d})^{2} + (c^{\pm1}_{fd})^{2} + (c^{\pm1}_{pd})^{2}\right)} 
\end{equation}
The expansion equation (\ref{eq:PAD(m=pm1)}) contains up to $P_6 (\cos\theta)$. The terms for $P_5 (\cos\theta)$ and $P_6 (\cos\theta)$ cancels with the corresponding terms in $I_0(\theta)$ (see below), so that $I(\theta)$ contains only up to $P_4 (\cos\theta)$.
The values of the amplitudes $c_d^{\pm 1}(=c_2^{\pm 1})$, $c_{pd}^{\pm 1}(=c_1^{\pm 1})$, and 
$c_{fd}^{\pm 1}(=c_3^{\pm 1})$ are already determined in section \ref{subsubsec:Partial wave amplitudes} (Table \ref{table:amp_TD-CASSCF}). 
We obtain $\beta_n\,(n=1,\cdots,6)$ values for $m=\pm{1}$ as a function of $\phi$ from TD-CASSCF simulations (markers in Fig.~\ref{fig:m1-fitting} and Table \ref{table:m1-fitting}).
Again, the odd-order parameters sinusoidally oscillate with $\phi$, while the even-order ones are constant.
Then, through global least-square fitting of equations (\ref{eq:beta1-m-pm1})-(\ref{eq:beta5-m-pm1}) to these TD-CASSCF results, we extract the values of $\Delta\eta_{pd}=\eta_{pd}-\eta_{d}$ and $\Delta\eta_{fd}=\eta_{fd}-\eta_{d}$.

The results are listed in Table \ref{table:scat_TD-CASSCF}.
The negligibly small standard errors and consistent values for the two simulation runs for $\hbar\omega=15.9$ eV (labels A and B) indicate successful fitting, which is also evident from the quasi-perfect agreement between the TD-CASSCF outputs and the reproduction from equations (\ref{eq:beta1-m-pm1})-(\ref{eq:beta6-m-pm1}) using the obtained amplitudes and phases in Fig.~\ref{fig:m1-fitting} and Table \ref{table:m1-fitting}.


\begin{table}[tb]
\caption{Comparison of $\beta_l\,(l=2,4,6)$, independent of $\phi$, for $m=\pm 1$ between the TD-CASSCF outputs and the reproduction by equations (\ref{eq:beta2-m-pm1}), (\ref{eq:beta4-m-pm1}), and (\ref{eq:beta6-m-pm1}) using the amplitudes and phases obtained through the global fitting procedure described in the text. The TD-CASSCF calculations were done at eight values of $\phi$ between 0 and $\frac{7}{4}\pi$ with an interval of $\frac{\pi}{4}$, and their average and standard deviation (stdev) are shown.}
\centering
\label{table:m1-fitting}
\begin{tabular}{cccccc}
\br
Label & $\hbar\omega$ (eV) & $\beta_l$ & \multicolumn{2}{c}{TD-CASSCF} & global-fitting \\
       &  &    &   average   &   stdev    &        \\
\mr
A & 14.3 & $\beta_2$ & $-0.967$ & $1.2\times 10^{-4}$ & $-0.968$  \\
  &      & $\beta_4$ & $1.48$ & $3.3\times 10^{-4}$ & $1.48$  \\
  &      & $\beta_6$ & $-1.51$ & $1.3\times 10^{-4}$ & $-1.51$  \\
B & 15.9 & $\beta_2$ & $-1.32$ & $2.3\times 10^{-4}$ & $-1.32$  \\
  &      & $\beta_4$ & $1.61$ & $2.8\times 10^{-5}$ & $1.61$  \\
  &      & $\beta_6$ & $-1.29$ & $2.4\times 10^{-4}$ & $-1.29$  \\
C & 15.9 & $\beta_2$ & $-1.24$ & $2.7\times 10^{-4}$ & $-1.24$  \\
  &      & $\beta_4$ & $1.47$ & $4.8\times 10^{-5}$ & $1.48$  \\
  &      & $\beta_6$ & $-1.24$ & $2.2\times 10^{-4}$ & $-1.24$  \\
D & 19.1 & $\beta_2$ & $0.617$ & $9.2\times 10^{-5}$ & $0.616$  \\
  &      & $\beta_4$ & $0.380$ & $1.6\times 10^{-4}$ & $0.378$  \\
  &      & $\beta_6$ & $-2.00$ & $8.2\times 10^{-5}$ & $-1.99$  \\    
\br
\end{tabular}
\end{table}

\begin{table}[tb]
\caption{$\Delta\eta_{fd}$, $\Delta\eta_{pd}$ and their standard errors extracted from $I_1(\theta)$ calculated with TD-CASSCF.}
\centering
\label{table:scat_TD-CASSCF}
\begin{tabular}{ccccccc}
\br
Label &  $\hbar\omega$ (eV) & \multicolumn{2}{c}{$\Delta\eta_{fd}$} & \multicolumn{2}{c}{$\Delta\eta_{pd}$} \\
&  & value     & error     & value     & error   \\
\mr
A & 14.3  & 1.144 & $5.6\times 10^{-3}$ & -2.353 & $7.2\times 10^{-3}$  \\
B & 15.9  &  1.184  & $2.8\times 10^{-3}$ & -2.269 & $3.2\times 10^{-3}$\\
C & 15.9  &  1.185 & $2.0\times 10^{-3}$ & -2.269 & $2.7\times 10^{-3}$ \\
D & 19.1  & 1.249 & $1.7\times 10^{-3}$ & -2.849  & $4.8\times 10^{-3}$ \\

\br
\end{tabular}
\end{table}

\subsubsection{Extraction of $\Delta\eta_{s}$, $\Delta\eta_{ps}$, $c_{pd}^0$, and $c_{ps}^0$ from $I_0(\theta)$ by global fitting.}

The PAD of $m = 0$ is expressed as,
\begin{eqnarray}
I_{0}(\theta)&=& 
|c_{s}^0 e^{i \left( \eta_{s} + \phi \right)} \, Y_{0}^{0}\left(\theta, \varphi \right) +
c_{ps}^0 e^{i \eta_{ps}} \, Y_{1}^{0}\left(\theta, \varphi \right) \nonumber \\
&+& c_{pd}^0 e^{i \eta_{pd}} \, Y_{1}^{0}\left(\theta, \varphi \right) +
c_{d}^0 e^{i \left( \eta_{d} + \phi \right)} \, Y_{2}^{0}\left(\theta, \varphi \right) \nonumber \\
&+& c_{fd}^0 e^{i \eta_{fd}} \, Y_{3}^{0}\left(\theta, \varphi \right) |^2 \label{eq:NePAD(m=-0)} \\
&=&\frac{B}{4\pi}\left[ 1+\sum _{ n=1 }^{ 6 }{ \beta _{ n } } { P }_{ n }(\cos { { \theta  }} )\right],
\end{eqnarray}
with,
\begin{equation}
    B = (c_d^0)^2+(c_{fd}^0)^2+2 c_{{pd}}^0 c_{{ps}}^0 \cos \left(\eta _{{pd}}-\eta _{{ps}}\right)+(c_{{pd}}^0)^2+(c_{{ps}}^0)^2+(c_s^0)^2,
\end{equation}
\begin{eqnarray}
    \label{eq:beta1-m-0}
    \beta_1 &=& \frac{1}{B}\left[\frac{18 c_d^0 c_{fd}^0 \cos \left(\eta _d-\eta _{fd}+\phi\right)}{\sqrt{35}}\right. \nonumber \\
    &+&4 \sqrt{\frac{3}{5}} c_d^0 c_{{pd}}^0 \cos \left(\eta_d-\eta_{{pd}}+\phi\right)+4 \sqrt{\frac{3}{5}} c_d^0 c_{{ps}}^0 \cos \left(\eta _d-\eta_{{ps}}+\phi\right) \nonumber \\  
    &+& \left. 2 \sqrt{3} c_s^0 c_{pd}^0 \cos \left(\eta_s-\eta_{pd}+\phi\right)+2 \sqrt{3} c_s^0 c_{ps}^0 \cos \left(\eta _s-\eta_{ps}+\phi\right) \right],
\end{eqnarray}
\begin{eqnarray}   
    \label{eq:beta2-m-0} 
    \beta_2 &=& \frac{1}{B}\left[ 2 \sqrt{5} c_d^0 c_s^0 \cos \left(\eta _d-\eta _s\right)+\frac{10 (c_d^0)^2}{7}  
    +6 \sqrt{\frac{3}{7}} c_{fd}^0 c_{{pd}}^0 \cos \left(\eta _{fd}-\eta _{{pd}}\right) \right. \nonumber\\
    &+& 6 \sqrt{\frac{3}{7}} c_{fd}^0 c_{{ps}}^0 \cos \left(\eta _{fd}-\eta _{{ps}}\right) +\frac{4 (c_{fd}^0)^2}{3} \nonumber\\
    &+& \left. 4 c_{{pd}}^0 c_{{ps}}^0 \cos \left(\eta _{{pd}}-\eta _{{ps}}\right)+2 (c_{{pd}}^0)^2+2 (c_{{ps}}^0)^2 \right]
\end{eqnarray}
\begin{eqnarray} 
    \label{eq:beta3-m-0}   
    \beta_3 &=& \frac{1}{B}\left[ \frac{8}{3} \sqrt{\frac{7}{5}} c_d^0 c_{fd}^0 \cos \left(\eta _d-\eta _{fd}+\phi\right)+6 \sqrt{\frac{3}{5}} c_d^0 c_{{pd}}^0 \cos \left(\eta _d-\eta _{{pd}}+\phi\right) \right. \nonumber \\  
    &+& \left. 6 \sqrt{\frac{3}{5}} c_d^0 c_{{ps}}^0 \cos \left(\eta _d-\eta _{{ps}}+\phi\right)+2 \sqrt{7} c_s^0 c_{fd}^0 \cos \left(\eta _s-\eta _{fd}+\phi\right)\right] 
\end{eqnarray} 
\begin{eqnarray} 
    \label{eq:beta4-m-0}     
    \beta_4 &=&\frac{2}{77B} \left[99 (c_d^0)^2+c_{fd}^0 \left\{44 \sqrt{21} \left[c_{{pd}}^0 \cos \left(\eta _{fd}-\eta _{{pd}}\right) \right.\right.\right. \nonumber \\  
    &+& \left.\left.\left. c_{{ps}}^0 \cos \left(\eta _{fd}-\eta _{{ps}}\right)\right]+63 c_{fd}^0\right\}\right] 
\end{eqnarray}
\begin{equation}    
    \label{eq:beta5-m-0}
    \beta_5 = \frac{20}{3B} \sqrt{\frac{5}{7}} c_d^0 c_{fd}^0 \cos \left(\eta _d-\eta _{fd}+\phi\right) 
\end{equation}
\begin{equation}   
    \label{eq:beta6-m-0}  
    \beta_6 = \frac{100 (c_{fd}^0)^2}{33B}
\end{equation}

\begin{figure}[tb]
\begin{center}
\includegraphics[width=10cm]{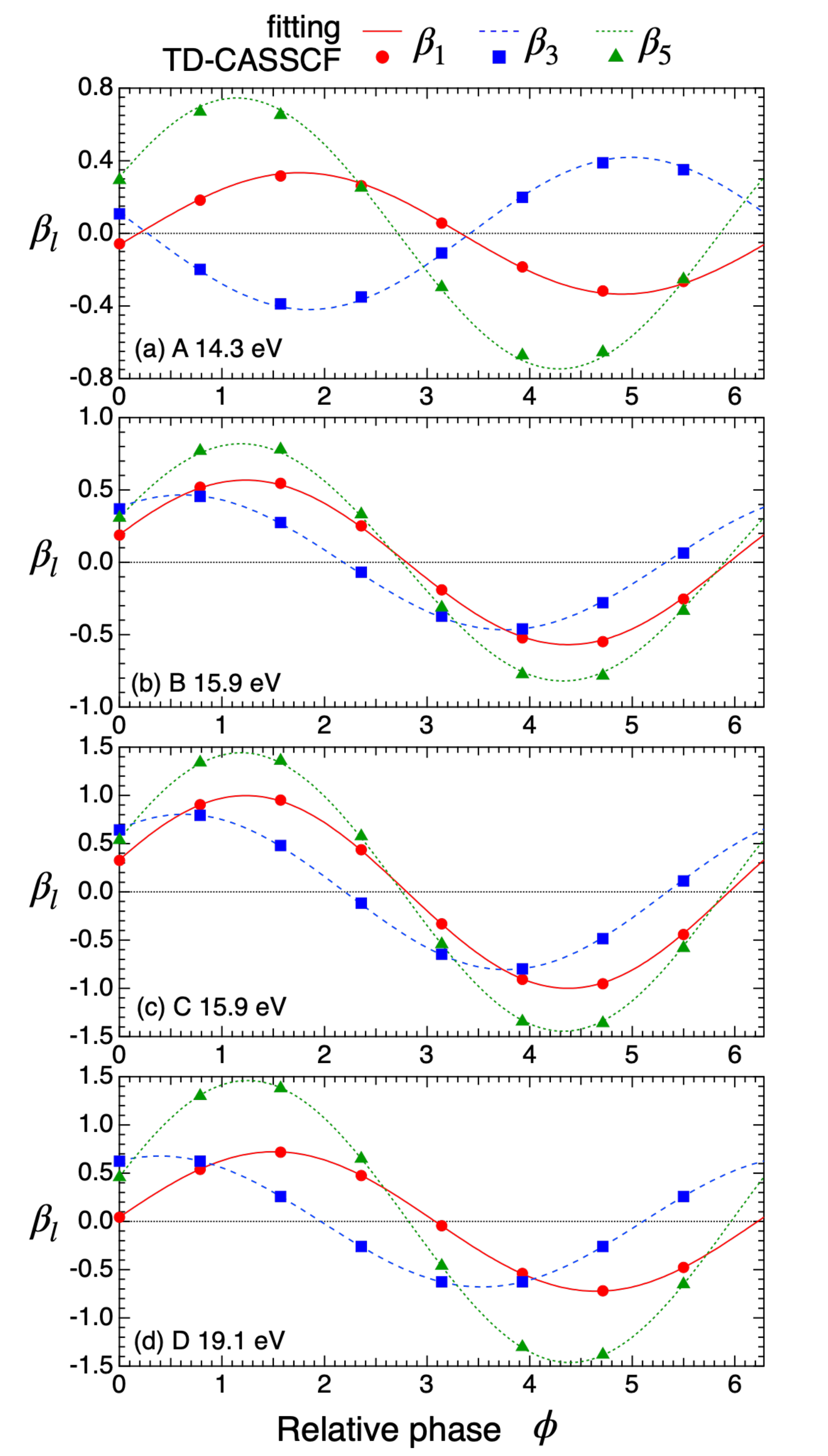}
\caption{Comparison of $\beta_l\,(l=1,3,5)$ as a function of $\phi$ for $m=0$ between the TD-CASSCF outputs (markers) and the reproduction by equations (\ref{eq:beta1-m-0}), (\ref{eq:beta3-m-0}), and (\ref{eq:beta5-m-0}) using the amplitudes and phases obtained through the global fitting procedure described in the text (lines). The TD-CASSCF calculations were done at eight values of $\phi$ between 0 and $\frac{7}{4}\pi$ with an interval of $\frac{\pi}{4}$.}
\label{fig:m0-fitting}
\end{center}
\end{figure}

We obtain $\beta_n\,(n=1,\cdots,6)$ values for $m=0$ as a function of $\phi$ from TD-CASSCF simulations (markers in Fig.~\ref{fig:m0-fitting} and Table \ref{table:m0-fitting}).
Using the already determined parameter values from Tables \ref{table:amp_TD-CASSCF} and \ref{table:scat_TD-CASSCF} as well as the relation $c_{pd}^0=\frac{4}{3}c_{pd}^{\pm 1} = \frac{4}{3}c_{1}^{\pm 1}$, further global fitting of these $\beta$ parameters determines the remaining parameters $\Delta\eta_{s}(=\eta_{s}-\eta_d)$,  $\Delta\eta_{ps}(=\eta_{ps}-\eta_d)$, $c_{pd}^0$, and $c_{ps}^0$ (Tables \ref{table:scat_TD-CASSCF_m_0} and \ref{table:scat_TD-CASSCF_m_0_ver2}).
Again, the standard errors are small, the two simulation runs for $\hbar\omega=15.9$ eV (labels A and B) deliver consistent results, and the fitting is nearly perfect (Fig.~\ref{fig:m0-fitting} and Table \ref{table:m0-fitting}).
Now, we have determined all the amplitude and phase values in equation (\ref{eq:NePADdecomposition}) as listed in Tables \ref{table:amp_TD-CASSCF}, \ref{table:scat_TD-CASSCF}, \ref{table:scat_TD-CASSCF_m_0}, and \ref{table:scat_TD-CASSCF_m_0_ver2}.

\subsection{Discussion}
\label{subsec:Discussion}

\begin{table}[tb]
\caption{Comparison of $\beta_l\,(l=2,4,6)$, independent of $\phi$, for $m=0$ between the TD-CASSCF outputs and the reproduction by equations (\ref{eq:beta2-m-0}), (\ref{eq:beta4-m-0}), and (\ref{eq:beta6-m-0}) using the amplitudes and phases obtained through the global fitting procedure described in the tex. The TD-CASSCF calculations were done at eight values of $\phi$ between 0 and $\frac{7}{4}\pi$ with an interval of $\frac{\pi}{4}$, and their average and standard deviation (stdev) are shown.}
\centering
\label{table:m0-fitting}
\begin{tabular}{cccccc}
\br
Label & $\hbar\omega$ (eV) & $\beta_l$ & \multicolumn{2}{c}{TD-CASSCF} & global-fitting \\
       &  &    &   average   &   stdev    &        \\
\mr
A & 14.3 & $\beta_2$ & $-0.280$ & $1.2\times 10^{-5}$ & $-0.280$  \\
  &      & $\beta_4$ & $-1.27$ & $5.4\times 10^{-4}$ & $-1.27$  \\
  &      & $\beta_6$ & $1.90$ & $2.3\times 10^{-4}$ & $1.89$  \\
B & 15.9 & $\beta_2$ & $2.01$ & $1.6\times 10^{-4}$ & $2.01$  \\
  &      & $\beta_4$ & $2.40$ & $2.7\times 10^{-4}$ & $2.39$  \\
  &      & $\beta_6$ & $2.68$ & $3.3\times 10^{-4}$ & $2.68$  \\
C & 15.9 & $\beta_2$ & $1.87$ & $2.8\times 10^{-4}$ & $1.87$  \\
  &      & $\beta_4$ & $2.36$ & $2.8\times 10^{-4}$ & $2.35$  \\
  &      & $\beta_6$ & $2.47$ & $6.7\times 10^{-4}$ & $2.47$  \\
D & 19.1 & $\beta_2$ & $0.629$ & $7.0\times 10^{-5}$ & $0.628$  \\
  &      & $\beta_4$ & $0.677$ & $8.5\times 10^{-5}$ & $0.678$  \\
  &      & $\beta_6$ & $2.58$ & $1.2\times 10^{-4}$ & $2.58$  \\    
\br
\end{tabular}
\end{table}

\begin{table}[tb]
\caption{$c_{pd}^0 (=\frac{4}{3}c_{pd}^{\pm 1} = \frac{4}{3}c_{1}^{\pm 1})$ as well as $c_{ps}^0$ and its standard error extracted from $I_0(\theta)$ calculated with TD-CASSCF.}
\centering
\label{table:scat_TD-CASSCF_m_0}
\begin{tabular}{ccccc}
\br
Label & $\hbar\omega$ (eV)  & $c_{pd}^0$ & \multicolumn{2}{c}{$c_{ps}^0$} \\
       &      &      &   value    & error       \\
\mr
A & 14.3      & $4.068\times 10^{-2}$ & $7.401\times 10^{-3}$ & $1.29\times 10^{-4}$  \\
B & 15.9      & $4.233\times 10^{-2}$ & $4.883\times 10^{-2}$ & $1.2\times 10^{-5}$ \\
C & 15.9      & $4.231\times 10^{-2}$ & $4.880\times 10^{-2}$ & $2.1\times 10^{-5}$ \\
D & 19.1      & $6.588\times 10^{-3}$ & $3.200\times 10^{-3}$ & $4.5\times 10^{-6}$ \\
\br
\end{tabular}
\end{table}

\begin{table}[tb]
\caption{ $\Delta\eta_{s}$, $\Delta\eta_{ps}$, and their standard errors extracted from $I_0(\theta)$ calculated with TD-CASSCF.}
\centering
\label{table:scat_TD-CASSCF_m_0_ver2}
\begin{tabular}{cccccc}
\br
Label & $\hbar\omega$ (eV)  & \multicolumn{2}{c}{$\Delta\eta_s$} & \multicolumn{2}{c}{$\Delta\eta_{ps}$} \\
& & value & error & value & error    \\
\mr
A & 14.3     & 2.270 & $1.2\times 10^{-2}$ & $-0.623$ & $9.5\times 10^{-3}$  \\
B & 15.9     & 2.014 & $1.2\times 10^{-2}$ & $1.135$  & $3.6\times 10^{-4}$ \\
C & 15.9     & 2.031 & $6.2\times 10^{-3}$ & $1.137$  & $7.1\times 10^{-4}$  \\
D & 19.1     & 1.756 & $8.0\times 10^{-4}$ & $-1.570$ & $3.8\times 10^{-3}$ \\
\br
\end{tabular}
\end{table}


In TD-MCSCF methods (including TDHF), the temporal evolution of the wave function is guided solely by the TDVP. 
The relations equation (\ref{eq:Wigner-Eckart}) are not a priori implemented.
How well they are satisfied serves as an estimate of numerical convergence with respect to the number of orbitals. 

As we have seen above, $c_\xi^1 = c_\xi^{-1}$ is satisfied by both TDHF and TD-CASSCF results.
$c_d^{\pm 1}/c_d^0 = \frac{\sqrt{3}}{2}=0.8660$ and $c_{fd}^{\pm 1}/c_{fd}^0=\frac{\sqrt{6}}{3}=0.8165$, especially the former, are satisfied better by TD-CASSCF than by TDHF (Table \ref{table:CG}).
For the case of TDHF, the difference from the theoretical values is larger for higher photon energy.
Indeed, while the PAD as a function of $\omega$-$2\omega$ relative phase $\phi$ obtained with the TDHF and TD-CASSCF methods agrees with each other at $\hbar\omega=14.3\,{\rm eV}$, the TDHF results deviate with increasing photon energy from the TD-CASSCF results, considered to be numerically more accurate \cite{You2020PRX}. 

Conditions B and C are common in photon energy $\hbar\omega$ and fundamental intensity $I_\omega$ but different in second-harmonic intensity $I_{2\omega}$.
We expect that the amplitudes $c_\xi^m (c_l^m)$ for two-photon pathways take the same values for both conditions, while those for single-photon ones scale as $\sqrt{I_{2\omega}}$.
These relations are well satisfied by the amplitudes calculated with the TDHF and TD-CASSCF methods (Tables \ref{table:amp_tdhf}, \ref{table:amp_TD-CASSCF}, \ref{table:scat_TD-CASSCF_m_0}, and \ref{table:amp_ratio})

Now that we have all the amplitude and phase values in equation (\ref{eq:NePADdecomposition}) listed in Tables \ref{table:amp_TD-CASSCF}-\ref{table:scat_TD-CASSCF_m_0_ver2}, we can easily calculate $\beta$ parameters (and PAD) for any combination of $\omega$ and $2\omega$ intensities as long as three or more photon ionization by $\omega$ and two or more photon ionization by $2\omega$ are negligible. This can be achieved simply by scaling $c_{ps}^0$, $c_{pd}^{0,\pm1}$, and $c_{fd}^{0,\pm1}$ as $I_\omega$ and $c_s^0$ and $c_d^{0,\pm1}$ as $\sqrt{I_{2\omega}}$.
Without the knowledge of the amplitudes and phases, since $\beta_l$'s $(l=1,\cdots,4)$ depend on $I_\omega$ and $I_{2\omega}$ in a complex, nonlinear manner, it would not be trivial, if not impossible, to scale the $\beta$ parameters calculated for one intensity pair to another, and it would be necessary to do simulations for each condition. 

\begin{table}[tb]
\caption{Ratios $c_{d}^1/c_{d}^0$ and $c_{fd}^1/c_{fd}^0$ from TDHF and TD-CASSCF simulations, compared with their theoretical values.}
    \centering
    \begin{tabular}{ccccc}
    \br
$\hbar\omega$ (eV)  & \multicolumn{2}{c}{$c_{d}^1/c_{d}^0$} & \multicolumn{2}{c}{$c_{fd}^1/c_{fd}^0$} \\
& TDHF & TD-CASSCF & TDHF & TD-CASSCF \\
    \mr
         14.3 & 0.8735 & 0.8560 & 0.8210 & 0.8232 \\
         15.9 & 0.8373 & 0.8604 & 0.8019 & 0.8333 \\ 
         15.9 & 0.8406 & 0.8521 & 0.8019 & 0.8334 \\ 
         19.1 & 0.8169 & 0.8641 & 0.7920 & 0.8099 \\
    \mr
         theory & \multicolumn{2}{c}{0.8660} & \multicolumn{2}{c}{0.816496} \\
    \br
    \end{tabular}
    \label{table:CG}
\end{table}

\begin{table}[tb]
\caption{The ratio between conditions C and B of $c_0^m$ ($c_s^m$) and $c_2^m$ ($c_d^m$), to be compared with $\sqrt{I_{2\omega}({\rm C})/I_{2\omega}({\rm B})}=1.89$.}
\centering
\label{table:amp_ratio}
\begin{tabular}{ccccccc}
\br
& \multicolumn{2}{c}{TDHF} & \multicolumn{2}{c}{TD-CASSCF} \\
	    $m$    & $c_0^m$ ($c_s^m$) & $c_2^m$ ($c_d^m$)  & $c_0^m$ ($c_s^m$)  & $c_2^m$ ($c_d^m$)  \\
\mr
$-1$ &      & 1.88 &      & 1.89 \\
$0$  & 1.88 & 1.88 & 1.89 & 1.91 \\
$1$  &      & 1.88 &      & 1.89 \\
\br
\end{tabular}
\end{table}

Let us finally emphasize the advantages of using coherent bichromatic setups rather than considering the single-photon ionization by $2\omega$ and the two-photon ionization by $\omega$ separately.
In the latter case, one could in principle extract the amplitudes as well as $\cos \left(\eta_{fd}-\eta_{pd}\right), \cos \left(\eta_{fd}-\eta_{ps}\right), \cos \left(\eta_{pd}-\eta_{ps}\right)$, and $\cos \left(\eta_{s}-\eta_{d}\right)$ using $\beta$ parameter values for $m=0$ and $\pm 1$. This approach, however, cannot distinguish $\left\{\eta_{fd}-\eta_{pd},\eta_{fd}-\eta_{ps},\eta_{pd}-\eta_{ps}\right\}=\{3.497,1.767,-1.730\}$ from $\{-3.497,-1.767,1.730\}$ and $\eta_{s}-\eta_{d}=2.270$ from $-2.270$, e.g., for $\hbar\omega=14.3\,{\rm eV}$.
In addition, the phases between the two-photon and single-photon paths such as $\eta_{fd}-\eta_{d}$ cannot be determined, either.
Our present method overcomes these problems, by examining how $\beta_l$'s for odd $l$, which describe $\omega$-$2\omega$ interference and are absent in the single-color cases, vary with $\phi$.
This aspect is indeed one of the virtues of coherent control. 

\section{Summary}
\label{sec:conclusions}

We have presented a successful evaluation of the amplitudes and phases of different photoionization paths from the TD-CASSCF simulation results for Ne irradiated by bichromatic XUV pulses.
On one hand, the amplitude $c_l^m$ of each partial wave is calculated as in equation (\ref{eq:partial wave amplitude}) during the tSURFF procedure for ARPES and PAD calculation.
The directly becomes the path amplitude $c_\xi^m$ if the single path leads to $(l,m)$. On the other hand, for the amplitudes of multiple paths resulting in the same final angular momenta ($p\to s \to p$ and $p\to d \to p$ in the present case) as well as the path phases $\eta_\xi$, we use global fitting of the $m$-resolved asymmetry parameters $\{\beta_l\}$ as a function of the $\omega$-$2\omega$ relative phase $\phi$, parametrized with $c_\xi^m$ and $\eta_\xi$, to the TD-CASSCF results.
By using the bichromatic setup, we can use $\beta_l$ with odd $l$, circumvent the 2-valuedness of arccos, and determine the phase difference between the SPI and TPI paths.

While we have presented the results for Ne with the TD-CASSCF methods, we can also treat other atomic systems.
Moreover, it will be straightforward to apply the present method in combination with various real-time {\it ab initio} approaches such as different types of TD-MCSCF methods with moving orbitals \cite{Zanghellini_2003,Kato_2004,Caillat_2005,Nguyen-Dang_2007,Miyagi_2013,Sato2013,Miyagi_2014,Haxton_2015}, the time-dependent configuration-interaction method \cite{Rohringer2006PRA,Greenman2010PRA,Sato2018AS,Teramura2019PRA}, the time-dependent optimized coupled cluster method \cite{Kvaal2012JCP,Sato2018JCP,Pathak2020JCP,Pathak2020JCPb,Pathak2020MP}, and the time-dependent density functional theory \cite{Gross1996TCC,Otobe2004PRA,Telnov2009PRA}.

\ack

We gratefully acknowledge support by the Cooperative Research Program of the Network Joint Research Center for Materials and Devices (Japan), the Dynamic Alliance for Open Innovation Bridging Human, Environment and Materials Program (Japan), Grant-in-Aid for Scientific Research (Grants No.~16H03881, No.~19H00869, and No.~JP19J12870) from the Ministry of Education, Culture, Sports, Science, and Technology of Japan (MEXT), JST COI (Grant No.~JPMJCE1313), and JST CREST (Grant No.~JPMJCR15N1).
K.U. acknowledges support via the X-ray Free Electron Laser Utilization Research Project and the X-ray Free Electron Laser Priority Strategy Program of MEXT, and the IMRAM Program of Tohoku University.
D.Y. acknowledges support by a Grant-in-Aid of Tohoku University Institute for Promoting Graduate Degree Programs Division for Interdisciplinary Advanced Research and Education.
\\

\appendix

\section{Relation of the complex partial wave amplitudes with $\{g_p^{l,m}\}$ and $\{D_q^p\}$}
\label{appendix:complex partial wave amplitudes}

In this appendix, we generalize equation (\ref{eq:partial wave amplitude}) to relate both the partial wave amplitudes and phases to $\{g_p^{l,m}\}$ and $\{D_q^p\}$.
While the photoelectron is in general an ensemble (incoherent mixture) of different coherent wave packets, let us assume that $\{g_p^{l,m}\}$ is obtained for one of them (e.g., each $m$ for the linear polarization case).
It follows from equations (\ref{eq:PEMD}) and (\ref{eq:apgY}) that,
\begin{equation}
	\rho ({\bf k})=\sum_{p,q}\sum_{l,l^\prime,m,m^\prime}g_p^{l,m}(k)Y_l^m(\theta_k,\varphi_k)g_q^{l^\prime,m^\prime}(k)^*Y_{l^\prime}^{m^\prime}(\theta_k,\varphi_k)^*D_q^p,
\end{equation}
and we rewrite it as,
\begin{equation}
	\label{eq:general1}
	\rho ({\bf k})=\sum_{l,l^\prime,m,m^\prime}\left[\sum_{p,q} g_p^{l,m}(k)g_q^{l^\prime,m^\prime}(k)^*D_q^p\right]Y_l^m(\theta_k,\varphi_k)Y_{l^\prime}^{m^\prime}(\theta_k,\varphi_k)^*.
\end{equation}
Here, without limiting ourselves to linear polarization, we take the mixing of different magnetic angular momenta into account.
%
Assuming the photoelectron wave packet is expressed as $\chi ({\bf k})=\sum_{l,m}b_l^m Y_l^m(\theta_k,\varphi_k)$ with $b_l^m=c_l^m e^{i\eta_l^m}$ being the complex amplitude, we have,
\begin{equation}
	\label{eq:general2}
	\rho ({\bf k})=\left|\chi ({\bf k})\right|^2=\sum_{l,l^\prime,m,m^\prime}b_l^m(k) b_{l^\prime}^{m^\prime *}(k) Y_l^m(\theta_k,\varphi_k)Y_{l^\prime}^{m^\prime}(\theta_k,\varphi_k)^*.
\end{equation}
Then, comparing equations (\ref{eq:general1}) and (\ref{eq:general2}), we find,
\begin{equation}
    \label{eq:photoelectron density matrix}
	b_l^m b_{l^\prime}^{m^\prime *} = \sum_{p,q} g_p^{l,m}\left(g_q^{l^\prime,m^\prime}\right)^*D_q^p.
\end{equation}
In principle, we can obtain $\{b_l^m\}$ from this (overdetermined) system of equations.
In mixed-state cases, equation (\ref{eq:photoelectron density matrix}) can be viewed as $(l,m)$-based photoelectron density matrix elements.
\\

\bibliographystyle{unsrt.bst}
\bibliography{jphysb}

\end{document}